\documentclass[12pt]{iopart}
\usepackage{iopams}  
\usepackage{epsf}  
\usepackage{epsfig}
\usepackage{float}  
\usepackage[]{cite}
\begin{document}

\title[Yang-Lee zeros for a nonequilibrium phase transition]
      {Yang-Lee zeros for a\\nonequilibrium phase transition}

\author{Stephan M Dammer\dag\footnote[2]{To whom correspondence 
should be addressed ({\tt dammer@comphys.uni-duisburg.de})}
, Silvio R Dahmen\S \\and
Haye Hinrichsen$\|$}

\address{\dag\ Theoretische Physik, Fakult\"at 4,
        Gerhard-Mercator-Universit{\"a}t Duisburg,
        47048~Duisburg, Germany}

\address{\S\ Instituto de Fisica, Universidade Federal do Rio Grande do Sul, 91501-970~Porto~Alegre~RS~Brazil}

\address{$\|$\ Theoretische Physik, Fachbereich 8, Universit{\"a}t
        Wuppertal,\\42097~Wuppertal, Germany}

\begin{abstract}
Equilibrium systems which exhibit a phase transition can be studied by
investigating the complex zeros of the partition function. This method,
pioneered by Yang and Lee, has been widely used in equilibrium
statistical physics. We show that an analogous treatment is possible for
a nonequilibrium phase transition into an absorbing state. By investigating the
complex zeros of the survival probability of directed percolation
processes we demonstrate that the zeros provide information about universal
properties. Moreover we identify certain non-trivial points where the survival probability for bond percolation 
can be computed exactly.  
\end{abstract}


%
%
\section{Introduction}\label{intro}
The investigation of nonequilibrium systems is of great importance
since most phenomena in nature take place under nonequilibrium
conditions. In this field, as in equilibrium physics,
phase transitions are particularly
interesting. However, when dealing with nonequilibrium systems one cannot
utilize such a well-established theoretical framework as in equilibrium statistical mechanics.
Therefore it is interesting to investigate which concepts of equilibrium physics can be transfered to nonequilibrium 
systems.
  
An important concept of equilibrium statistical physics is the theory of Yang and Lee~\cite{YangLee}
for the emergence of nonanalytic behaviour at phase transtitions.
Only recently these ideas have been applied to an integrable
nonequilibrium model~\cite{Arndt}. In the present work we use similar techniques to investigate 
directed percolation (DP)~\cite{Physica} as a paradigm for {\em nonintegrable} systems far from equilibrium. 
We show that it is possible to transfer the ideas of Yang and Lee to DP and that
this method does indeed provide information about universal
properties of the phase transition.  
\subsection{Yang and Lee's theory in equilibrium statistical physics}\label{YL_equi}
We shall shortly sketch the ideas of complex zeros of the partition
function in equilibrium statistical mechanics. Taking the Ising
model without an external magnetic field as an example,
we consider a system of $N$ spins $s_i$ ($s_i=\pm 1,\, i=1,\ldots
,N$) on a lattice where the spins interact with their nearest
neighbours, denoted by $<i,j>$. The total energy ${\cal
  H}(\left\{s_i\right\})$ of a configuration $\left\{s_i\right\}$ of
the spins is given by
${\cal H}(\left\{s_i\right\})=-J\sum _{<i,j>}s_is_j$.
The system is said to be in thermal equilibrium at temperature $T$
when the probability distribution
$P(\left\{s_i\right\})$ to find the system in configuration
$\left\{s_i\right\}$ is given by the stationary Gibbs ensemble $P(\left\{s_i\right\})\propto \rme^{-\beta {\cal
    H}(\left\{s_i\right\})},\ \beta = k_{\rm B}T$, where $k_{\rm B}$ is Boltzmann's
constant. In that case the statistical properties of the system are
fully determined by the canonical partition function
\begin{eqnarray}
Z_N(T)&=&\sum _{\left\{s_i\right\}}\rme^{-\beta {\cal
    H}(\left\{s_i\right\})}=\sum
    _{\left\{s_i\right\}}x^{r(\left\{s_i\right\})}\quad {\rm with}\label{Zustandssumme}\\\nonumber\\
&&x=\rme^{\beta J}>0\quad , \quad r(\left\{s_i\right\})=\sum
    _{<i,j>}s_is_j\ .\label{x_und_r}
\end{eqnarray} 
For finite system size $Z_N$ can be expressed as the ratio of two
polynomials in $x$ with integer coefficients.
The free energy per lattice site is obtained from $Z_N$ through
\begin{eqnarray}\label{f_N}
f=\lim_{N\to\infty}\left( -\frac{1}{\beta N}\ln Z_N\right)
\end{eqnarray}
The hallmark of phase transitions is the appearance of
nonanalyticities. Since in Equation~(\ref{f_N}) singularities occur as
zeros of the partition function $Z_N$, these are good candidates for
indicating a phase transition. However, $Z_N$ has no zeros
for a positive real temperature so that there is no phase transition for finite $N$. This led Yang and Lee to study
the complex zeros of $Z_N$\footnote[5]{For a field-driven
  phase transition these zeros are denoted as {\em Yang-Lee zeros} while
  in the case of a temperature-driven phase transition the term {\em
    Fisher zeros} is used.}, which in this case correspond to complex temperatures, and examine their behaviour as the system size
grows. They argued that in the thermodynamic limit $N\to\infty$ in
(\ref{f_N}), if the model exhibits a phase transition, an ever increasing number of complex zeros accumulate
in the vicinity of a point on the positive real axis, which corresponds to a physical temperature. The zeros
thereby induce singularities at the critical point in
the thermodynamic limit, thus explaining the crossover to nonanalyticity
at the transition in the limit of an infinite system size. Such a
behaviour has been observed for many equilibrium models.
Moreover, several features of the distribution of zeros have been
related to universal properties of the system under
consideration~\cite{YangLeearticles}.      
\subsection{Directed percolation, a process away from equilibrium}
Since nonequilibrium systems do not obey the
stationary Gibbs ensemble they cannot be handled using the partition function of
equilibrium statistical physics. Instead they are described by the time-dependent
probability distribution $P_t(\left\{s_i\right\})$ which has to be
derived from the master equation. In most cases this cannot be done
exactly. Nevertheless, the concept of universality, well-known from
equilibrium statistical physics, proves to be suitable for
nonequilibrium phase transitions as well. In this context models exhibiting a transition from
a fluctuating active phase to a nonfluctuating inactive phase
(absorbing state) have been extensively studied~\cite{Kinzel,MarroDickman,Hinrichsen1,Hinrichsen2}. 
These models are used to describe spreading processes as e.g. forest
fires, where a spreading agent can either spread over the entire system or die
out after some time. The two phases are seperated by a nonequilibrium
phase transition. The most prominent universality class of transitions
into an absorbing state is that of directed percolation. Directed
percolation is an anisotropic variant of ordinary
percolation in which activity can only percolate along a given
direction in space. Regarding this direction as a temporal
degree of freedom, DP can be interpreted as a dynamical process. Directed
percolation emerges in a variety of physical problems ranging from catalytic reactions on 
surfaces~\cite{Ziff} to spatio-temporal intermittency in magnetic fluids~\cite{Bayreuth}.   
\subsection{Realizations of DP}\label{dpreal}
\begin{figure}
\epsfxsize=74mm
\centerline{\epsffile{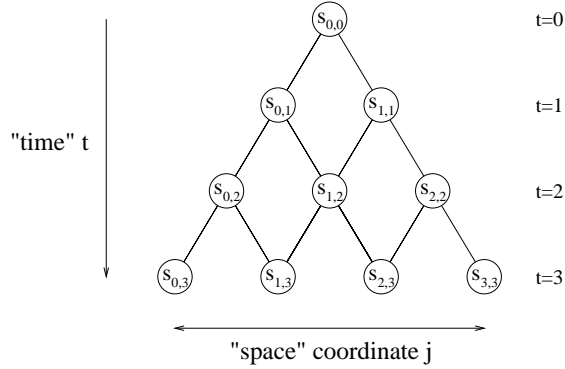}}
\caption{\label{FigBondDP}
Lattice geometry of DP on a tilted square lattice in 1+1 dimensions. In both
cases (DPb and DPs) sites can be eiter active ($s_{j,t}=1$) or
inactive ($s_{j,t}=0$). In DPb activity can percolate
forward in time through bonds (solid lines) which are conducting. In DPs
acitvity can percolate through bonds that connect permeable sites.}
\end{figure}

Simple realizations of DP are directed bond (DPb) and directed site
percolation (DPs) on a tilted square lattice (see Figure~\ref{FigBondDP}). In
directed bond percolation the bonds are conducting with probability $p$ and
non-conducting with probability $1-p$. In this model
sites at time $t>0$ are activated by directed paths of conducting bonds, originating from active
sites at time $t=0$. A cluster consists of all sites that are connected by such paths of
conducting bonds to active sites in the initial state. In directed site percolation on the other hand
all bonds are conducting while the sites themselves can be either permeable~($p$) or blocked~($1-p$). Activity can 
spread from permeable site to permeable
site. A cluster is formed by permeable sites that are connected to active sites at time
$t=0$ by a directed path of bonds that only connect permeable sites. 

The order parameter which characterizes the phase
transition is the probability $P(\infty)$ that a randomly chosen site
belongs to an infinite cluster. For $p>p_{\rm c}$ this probability is
finite whereas it vanishes for $p \leq p_{\rm c}$. Close to the phase
transition $P(\infty)$ is known to vanish algebraically as $P(\infty) \sim (p-p_{\rm
  c})^\beta$, where $\beta$ is an universal critical exponent. In addition, the DP process is
characterized by a spatial correlation length $\xi _{\bot}$
(perpendicular to time) and a temporal correlation length $\xi
_{||}$. As $p$ approaches $p_{\rm c}$ these length scales
are known to diverge as
\begin{eqnarray}
\xi _{\bot}\sim |p-p_{\rm c}|^{-\nu _{\bot}}\ ,\qquad \xi _{||}\sim |p-p_{\rm c}|^{-\nu _{||}}\label{xi}
\end{eqnarray}
with the critical exponents $\nu _{\bot}$ and $\nu _{||}$. The scaling
behaviour (\ref{xi}) implies that DP is invariant under the scaling transformations
\begin{eqnarray}
x\to\Lambda x\ ,\quad t\to\Lambda ^z t\ ,\quad (p-p_{\rm c})\to\Lambda
^{-1/\nu _{\bot}}(p-p_{\rm c})\ ,
\end{eqnarray}
where $x$ is the position and $z=\nu _{||}/\nu _{\bot}$ is
the so-called dynamic exponent. Numerical estimates for the values of
$p_{\rm c}$ and the critical exponents in 1+1 dimensions are $p_{\rm
  c}=0.6447001(1)$, $\beta=0.27649(4)$, $\nu
_{\bot}=1.096844(14)$ and $\nu _{||}=1.733825(15)$~\cite{Jensen} (bond percolation) and $p_{\rm
  c}=0.705489(4)$~\cite{Yu} (site percolation).  

Although DP can be defined and easily simulated, it is one of
the very few systems for which -- even in one spatial dimension -- no
analytical solution is known, suggesting that DP is a non-integrable
process. In fact, the values of the percolation threshold and the critical
exponents are not simple numerical fractions but seem to be irrational
instead. 
\subsection{Yang and Lee zeros for directed percolation}
To apply the idea of Yang-Lee zeros to DP, we consider the order parameter
in a {\em finite} system as a function of the percolation probability $p$ in the
complex plane. This can be done by studying the time-dependent survival probability
$P(t)$, which is defined as the probability that a cluster generated in a
single site at time $t=0$ survives up to time $t$ (or even
longer). For a finite system $P(t)$ can be expressed as a polynomial
in $p$ (see below). Note that $\lim_{t\to\infty}P(t)$ and the order
parameter $P(\infty)$ coincide. 

The survival probability and the
partition function (\ref{Zustandssumme}) show a similar behaviour in
many respects. For finite systems $P(t)$ and $Z_N$ do not have
relevant zeros in the physical region of the control parameter $0\leq
p\leq 1$ and $0<T<\infty $ although the phase transition is
marked by a vanishing $P(t)$ and $Z_N$ at the critical points in the
limit of infinite systems. At those points both functions exhibit
nonanalytic behaviour. The zeros of $P(t)$ and $Z_N$ are generated
by polynomials with real integer coefficients, thus the zeros come in complex-conjugate
pairs. 

According to Yang and Lee the increasing system size is accompanied by
an approach of some of the complex zeros of $Z_N$ to the physical region of
the control parameter, while the accumulation point marks the critical
point. We show that the complex zeros of $P(t)$ approach the
critical value $p_{\rm c}$ on 'trajectories' for increasing time $t$
and that the distance between the zeros and the critical point is
related to the critical exponent for the temporal correlation length. 
\subsection{Determining the survival probability}
\begin{figure}
\epsfxsize=95mm
\centerline{\epsffile{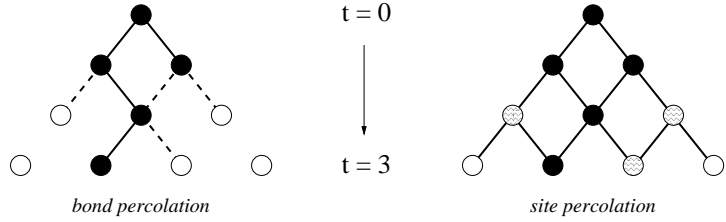}}
\caption{\label{FigCluster}
Example of clusters which survive until time $t=3$. Acitve sites are shown as
black circles, conducting bonds as solid lines. Left: Dashed lines denote the
hull, irrelevant bonds are not shown. The corresponding weight to the
survival probability $P(3)$ in~(\ref{SurvivalProbability}) is
$p^4(1-p)^4$. Right: White patterned circles belong to the hull. Simple white
circles are irrelevant for the survival of the cluster. The corresponding
weight to $P(3)$ is $p^4(1-p)^3$.}  
\end{figure}

In directed bond (site) percolation the survival probability $P(t)$ is given
by the sum over the weights of all possible configurations of bonds
(sites) for which the process survives at least up to time $t$. Each
conducting bond (permeable site) contributes to the weight with a factor
$p$, while each non-conducting bond (blocked site) contributes with a
factor $1-p$. However, the states of those bonds (sites) which do
not touch the actual cluster are irrelevant as they do not contribute to the
survival of the cluster. Therefore, it is sufficient to consider the
sum over all possible clusters {$\cal C$} of bonds (sites) connected to
the origin. Each cluster is weighted by the contributions of the
conducting bonds (permeable sites) belonging to the cluster and the
non-conducting bonds (blocked sites) belonging to its hull. Roughly speaking, the hull surrounds
the cluster. More precisely the hull of a cluster is the set of non-conducting bonds (blocked sites) that would
contribute to the cluster if they were conducting (permeable) (see
Figure~\ref{FigCluster}). Thus the survival probability can be
expressed as
\begin{equation}
\label{SurvivalProbability}
P(t) = \sum_{\cal C} p^n (1-p)^m \,,
\end{equation}
where the sum runs over all clusters reaching the horizontal row at time $t$.
For each cluster $n$ denotes the number of its bonds (sites), while $m$ is the
number of bonds (sites) belonging to its hull. Note that in this sense the hull does not include bonds
connecting sites at time $t$ and $t+1$ (DPb) or sites at time $t+1$ (DPs) since
the cluster may survive even longer. Summing up all weights in
Equation~(\ref{SurvivalProbability}) one obtains a polynomial. As can be
verified, the first few polynomials for the survival probability in the
directed bond percolation process are 
\newpage
\begin{eqnarray}\label{polynomials}
P(0)=1\\
P(1)=2p-p^2 \nonumber \\         
P(2)=4p^2-2p^3-4p^4+4p^5-p^6 \nonumber \\
P(3)=8p^3-4p^4-10p^5-3p^6+18p^7+5p^8-30p^9+24p^{10}-8p^{11}+p^{12}
\nonumber \\ P(4)=16p^4-8p^5-24p^6-8p^7+6p^8+84p^9-29p^{10}-62p^{11}-120p^{12} \nonumber \\
\qquad\quad+244p^{13}+75p^{14}-470p^{15}+495p^{16}-268p^{17}+83p^{18}
-14p^{19}+p^{20}\,. \nonumber
\end{eqnarray}
As $t$ increases the number of cluster configurations grows rapidly,
leading to complicated polynomials with very large coefficients (e.g.
for $t=15$ the largest coefficient for bond percolation is of order $10^{44}$). 
\\

The paper is structured as follows. In Section~\ref{universality} we investigate if the
distribution of zeros is totally dependent on microscopic details of
the underlying model or whether it also contains universal
features. We ask whether the distribution is related to some of
the critical exponents which characterize the phase transition of
directed percolation. Section~\ref{Exact} presents certain non-trivial
points where the polynomials $P(t)$ for bond percolation can be
calculated exactly for all times. Section~\ref{conclusion} ends with a
conclusion. In \ref{MC} we show that the value of the survival probability
can be computed by Monte Carlo simulations even 
if $p$ is a complex number. The first coefficients of the polynomials
$P(t)$ for bond percolation are discussed in \ref{coefficients}.
\section{Universal properties}\label{universality}
\begin{figure}[b]
\epsfig{file=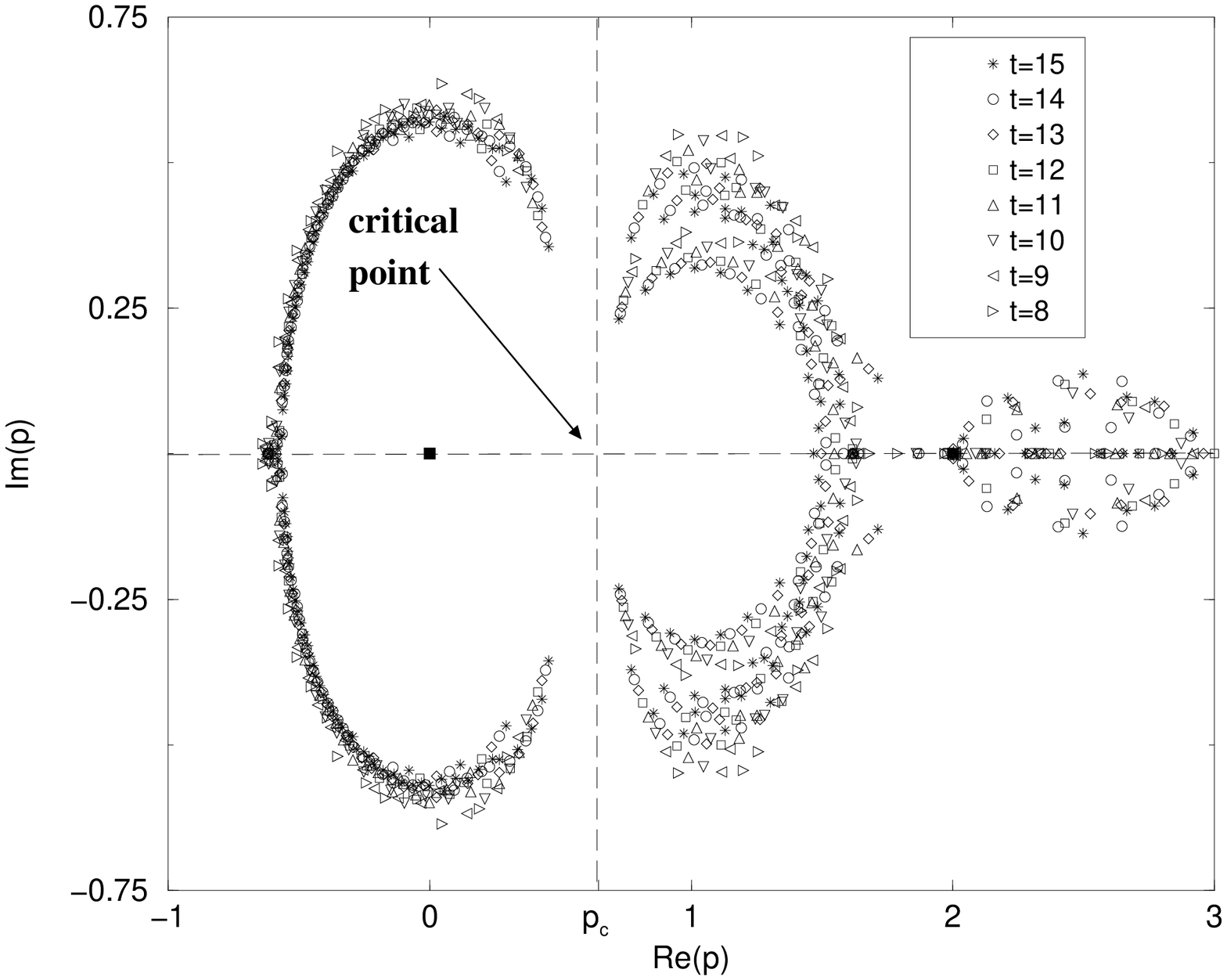,width=0.37\linewidth,angle=270}
\hspace*{5mm} 
\epsfig{file=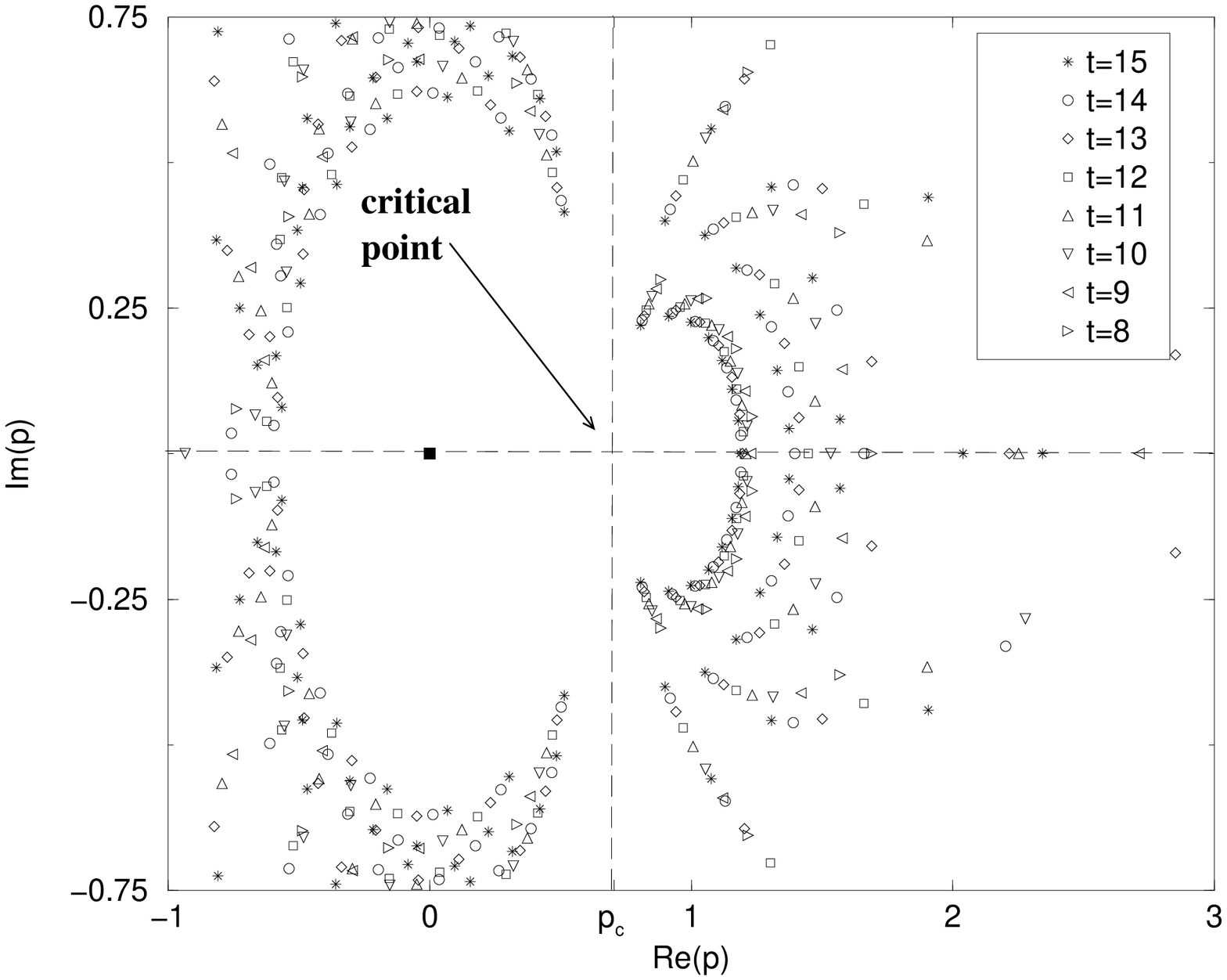,width=0.37\linewidth,angle=270}
\caption{\label{zerosshow} Zeros of the survival probability
  $P(t)$ in the complex plane.\\Left: Directed bond percolation. Right:
  Directed site percolation.}  
\end{figure}

\begin{figure}[t]
\epsfig{file=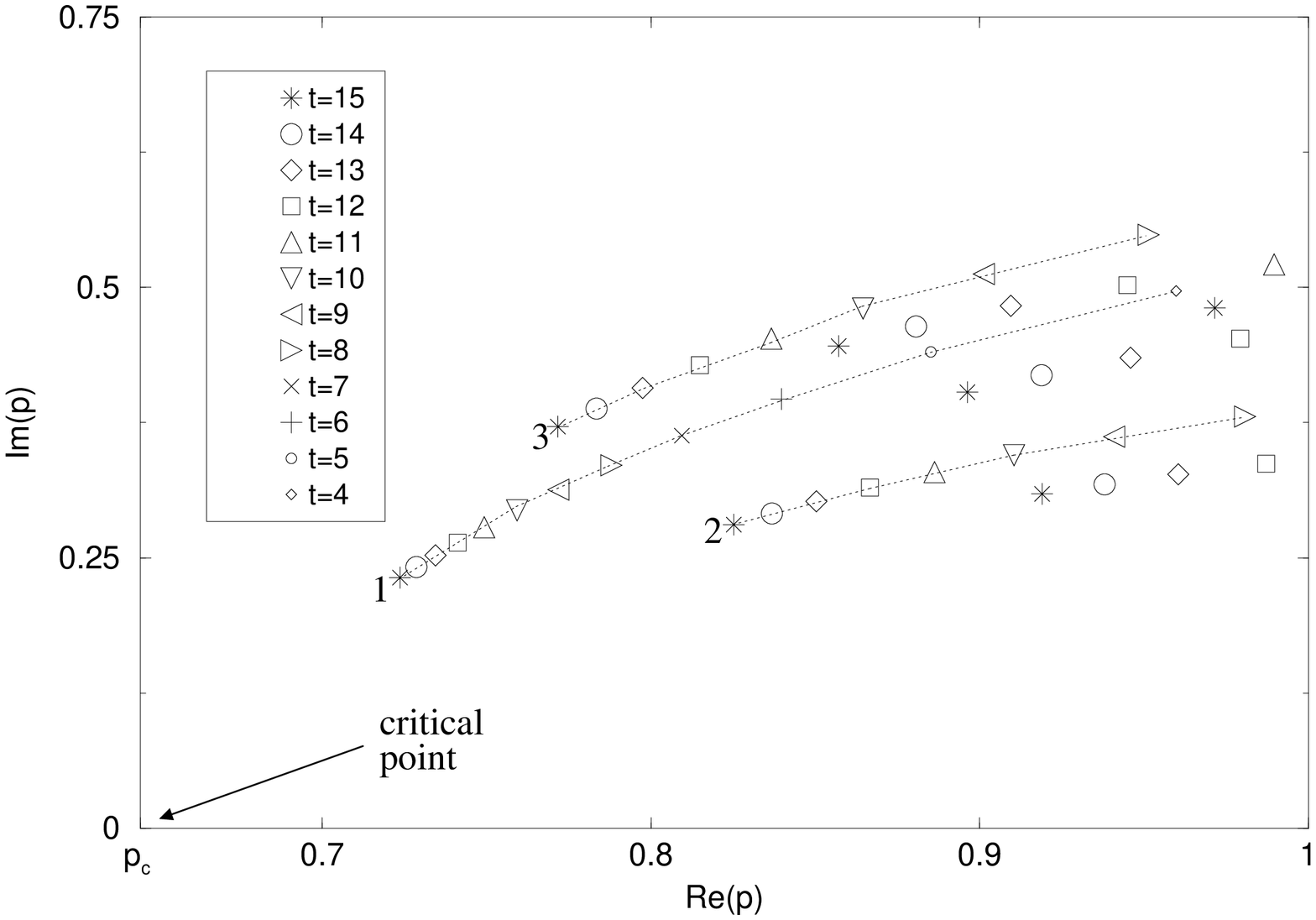,width=0.37\linewidth,angle=270}
\hspace*{5mm}
\epsfig{file=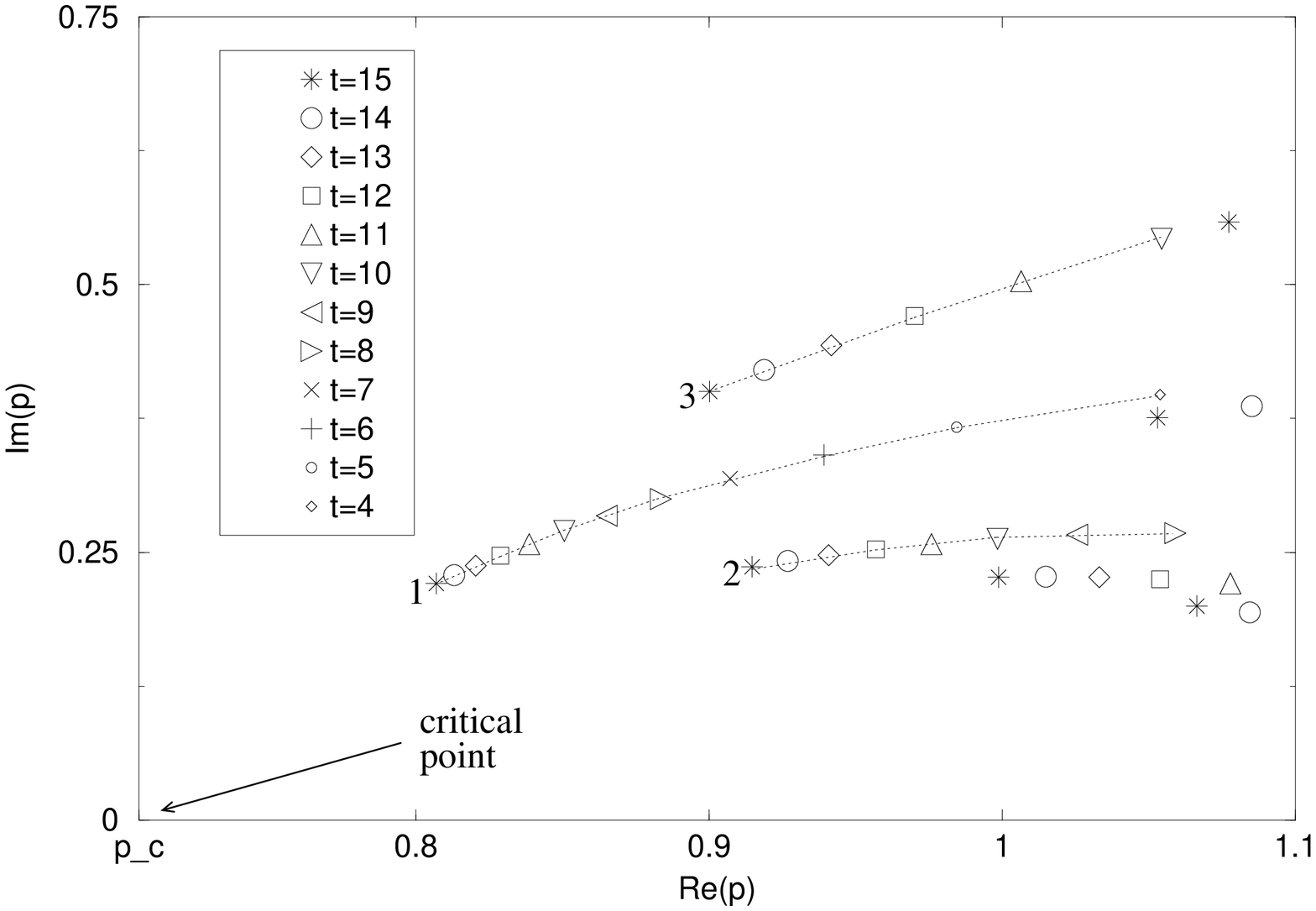,width=0.37\linewidth,angle=270}
\caption{\label{trajshow} Zeros approaching the critical
  point. Three `trajectories' are enumerated and shown as dashed lines. Left: Directed
  bond percolation. Right: Directed site percolation.}  
\end{figure}

The distribution of the zeros of $P(t)$ (from $t=8$ to $t=15$) in the neighbourhood of the critical
point is shown in Figure~\ref{zerosshow}.
Away from the critical point the appearance of the distributions for
DPb and DPs is quite different. This implies that in this region the distribution of
zeros depends strongly on microscopic details and hence is non-universal. However, there is also a
general feature which can be observed in both cases. The innermost zeros
approach the critical point on `trajectories' as $t$ increases (see Figure~\ref{trajshow}).
 We applied a standard Bulirsch-Stoer (BST)
acceleration algorithm~\cite{henkelandschuetz} to the set of zeros of each enumerated `trajectory'
of Figure~\ref{trajshow} to determine $\lim_{t\rightarrow\infty} {\rm Re}\left(p^{\rm zero}(t)\right)$ and 
$\lim_{t\rightarrow\infty} {\rm Im}\left(p^{\rm zero}(t)\right)$. The results are
listed in Table \ref{p_c_extrapo}.
\begin{table}[b]
\center
\renewcommand{\arraystretch}{1.2}
\begin{tabular}{l|cc||l|cc}\hline\hline
trajectory &  ${\rm Re}\left(p^{\rm zero}_{t\to\infty}\right)$ & ${\rm
  Im}\left(p^{\rm zero}_{t\to\infty}\right)$ & trajectory &  ${\rm
  Re}\left(p^{\rm zero}_{t\to\infty}\right)$ & ${\rm Im}\left(p^{\rm zero}_{t\to\infty}\right)$\\\hline\hline
bond 1 & 0.64472(1) & 0.0001(1) & site 1 & 0.70547(3) & 0.0001(3)\\
bond 2 & 0.6445(2) & 0.008(1) & site 2 & 0.709(4) & 0.006(1)\\
bond 3 & 0.6470(4) & 0.051(7) & site 3 & 0.712(1) & 0.01(1)\\\hline\hline
\end{tabular}
\caption{ Bulirsch-Stoer extrapolants for the zeros of the
  `trajectories' from Figure \ref{trajshow} for both
bond and site percolation in 1+1 dimensions.\label{p_c_extrapo} }
\end{table}
Although we calculated the zeros only for small systems (until $t=15$)
the extrapolants accord fairly well with the numerical values of the percolation
threshold\footnote[6]{The convergence for site percolation is slower than
for bond percolation since the order of the polynomials $P(t)$
for DPs is smaller than for DPb.} $p_{\rm c}=0.6447001(1)$ (bond percolation) and $p_{\rm
  c}=0.705489(4)$ (site percolation).    
Thus a similar scenario as observed by Yang and Lee for equilibrium phase
transitions proves to be suitable for the nonequilibrium phase
transition of DP. As time increases the zeros of the survival probability
$P(t)$ approach the real axis between $p=0$ and $p=1$. The accumulation
point is the critical point. 

\begin{figure}[h]
\epsfig{file=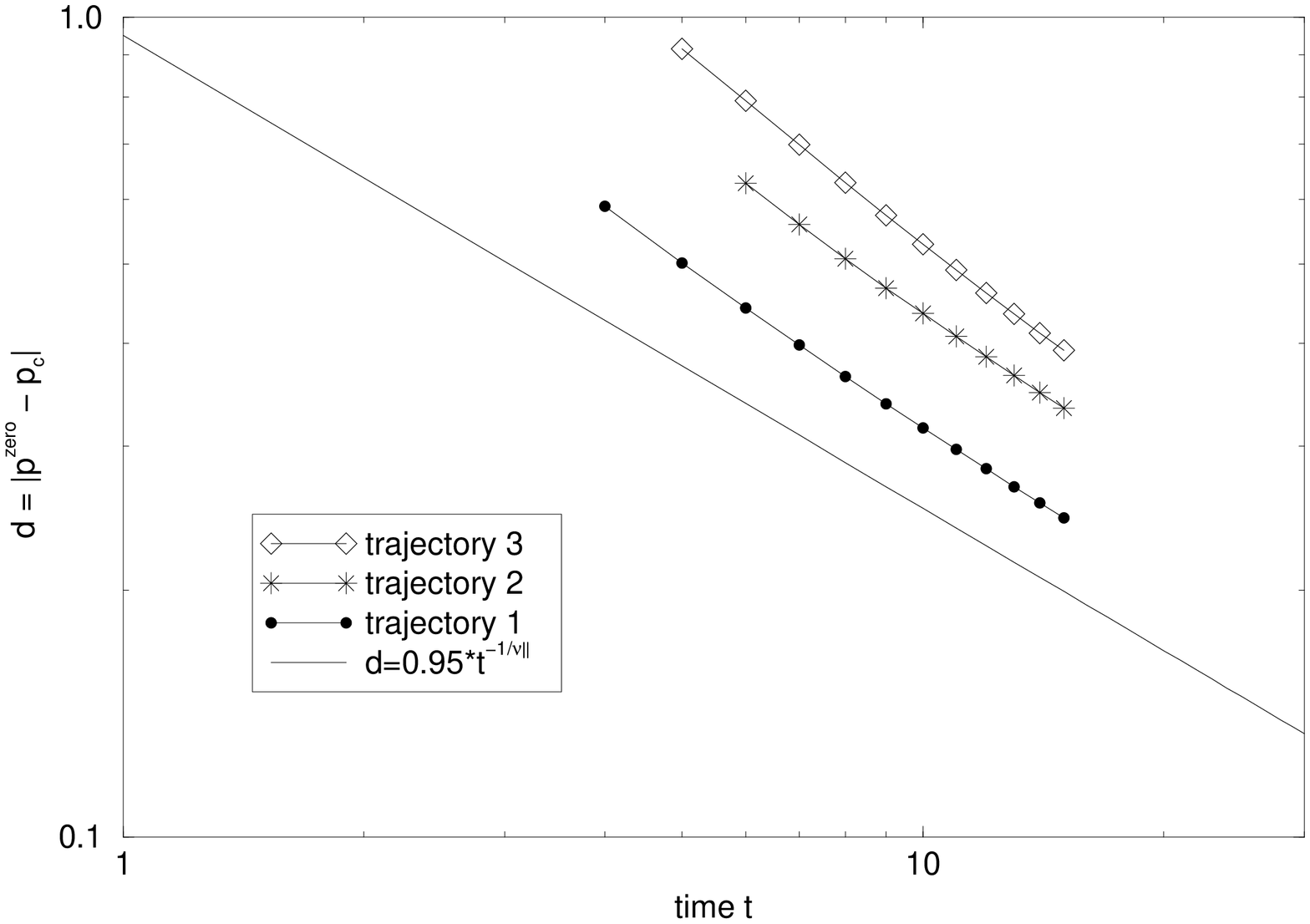,width=0.37\linewidth,angle=270}
\hspace*{5mm}
\epsfig{file=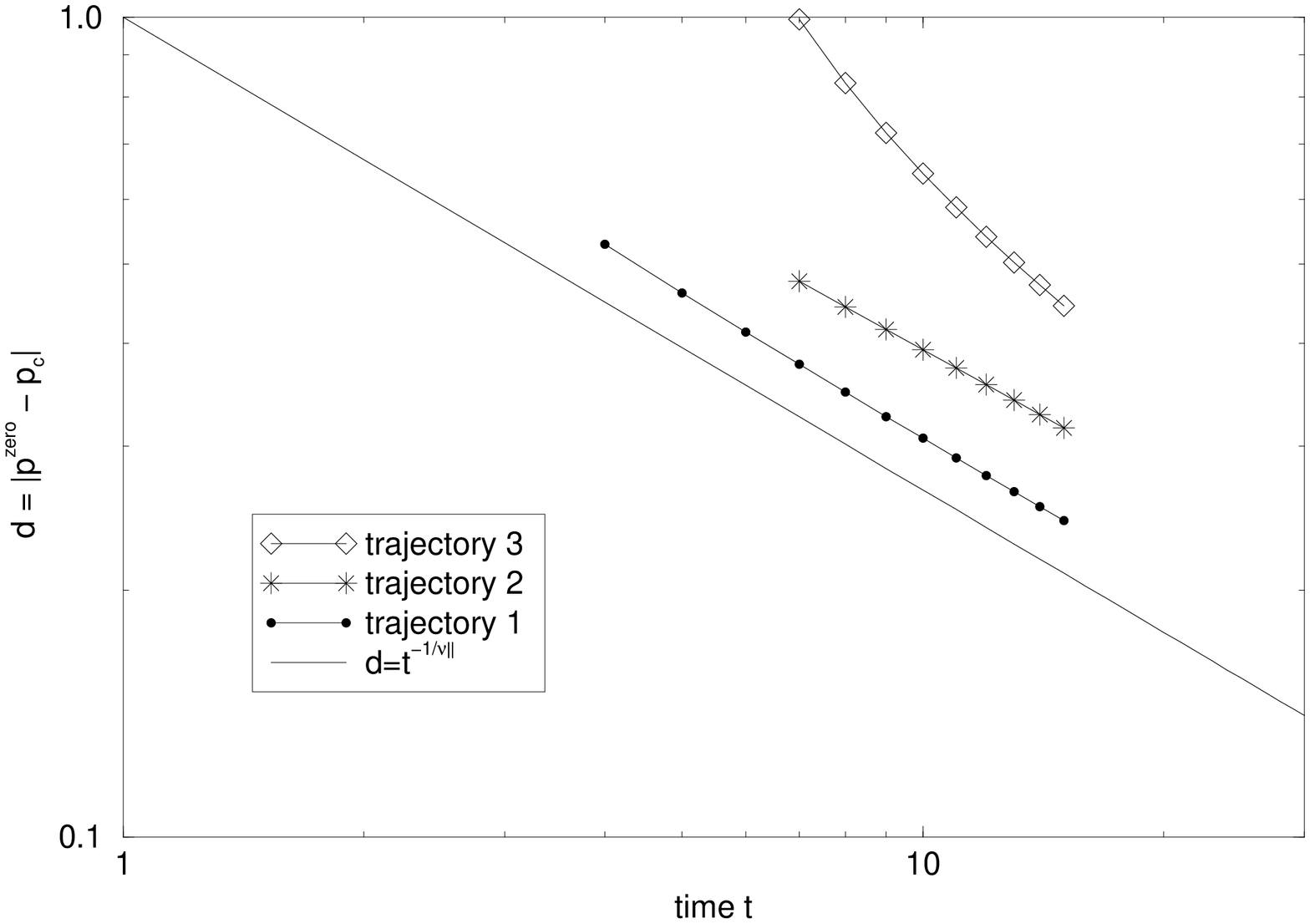,width=0.37\linewidth,angle=270}
\caption{\label{trajexp}The distance from the critical point $d=|p^{\rm
    zero}-p_{\rm c}|$ versus time $t$. For comparison the power law
    $d\sim t^{-1/\nu _{||}}$ is shown, with $\nu _{||}=1.7338$. Left: Directed
  bond percolation. Right: Directed site percolation.}  
\end{figure}

So far we have shown that the zeros of $P(t)$ provide information
about the existence of the phase transition and about the critical
value $p_{\rm c}$. However, the value of $p_{\rm c}$ is
non-universal and depends on the particular realization of
DP. On the contrary universal features are independent of the microscopic
details of the underlying process. On each
`trajectory' of Figure~\ref{trajshow} we
calculated the distance between the zeros and the critical
point $d(t)=|p^{\rm zero}(t)-p_c|$ with the values of $p_{\rm c}$
as given in section \ref{dpreal}. According to Equation (\ref{xi}) and
simple scaling arguments we expect $d(t)$ to decrease as $d(t)\sim
t^{-1/\nu_{||}}$. Application of the BST algorithm yields results which
support this claim, as shown in Table \ref{nu_extrapo}.
\begin{table}[t]
\center
\renewcommand{\arraystretch}{1.2}
\begin{tabular}{l|c||l|c}\hline\hline
trajectory &  $1/\nu_{||}$  & trajectory & $1/\nu_{||}$\\\hline\hline
bond 1 & 0.57675(3) & site 1 & 0.5765(7)\\
bond 2 & 0.575(2) & site 2 & 0.5771(2)\\
bond 3 & 0.576(4) & site 3 & 0.5731(2)\\\hline\hline
\end{tabular}
\caption{ Bulirsch-Stoer extrapolants for the exponent $1/\nu_{||}$. On
  each `trajectory' in Figure \ref{trajshow} the distance
  $d(t)=|p^{\rm zero}(t)-p_c|$ is assumed to decrease as $d(t)\sim
  t^{-1/\nu_{||}}$. \label{nu_extrapo} }
\end{table}
The extrapolants are in fairly good agreement with the numerical value
of $1/\nu_{||}=0.57676$, even for
small systems. This means that universal properties of the phase transition of
directed percolation are encoded in the complex zeros of the survival
probability $P(t)$. Figure \ref{trajexp}
shows the used data and the asymptotic power law $d(t)\sim
t^{-1/\nu_{||}}$ with the numerical value of $\nu_{||}$.    
\section{Exact results}\label{Exact}
In this section we will address a particularly surprising
observation, namely the existence of certain points on the real axis
where the polynomials $P(t)$ for bond percolation can be solved
exactly for all values of $t$. Beside the trivial points $p=0$ (where $P(t)=\delta_{t,0}$) and
$p=1$ (where $P(t)=1$) we find a $t$-independent zero at $p=2$ and,
even more surprisingly, a very simple solution if $p$ is equal to one of the Golden Ratios
$(1\pm \sqrt{5})/2$. The Golden Ratios are the roots of the quadratic equation
$p^2=p+1$ and play an important role not only in number
theory~\cite{NumberTheory} but also in other fields ranging from chaotic
systems~\cite{chaotic} to arts~\cite{arts}. Although these special
points are located outside the physically accessible region $0 \leq p
\leq 1$, their existence may be helpful for further investigations of
the polynomials $P(t)$. 
\subsection{Time-independent zero at $p=2$}
\begin{figure}[t]
\epsfxsize=80mm
\centerline{\epsffile{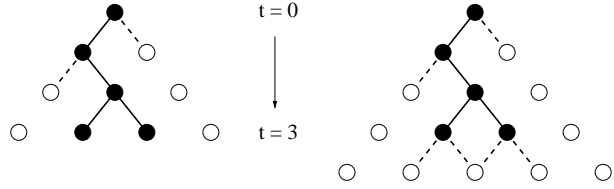}}
\caption{\label{FigCluster2} 
Example for different weights for $P(t)$ and $R(t)$. Left: The contribution of
the shown cluster to $P(3)$ in~(\ref{SurvivalProbability}) is
$p^4(1-p)^2$. Right: The contribution of the cluster to $R(3)$
in~(\ref{DieoutProbability}) is $p^4(1-p)^6$ since it is required that active
sites at $t=3$ are not connected to sites at $t=4$ by active bonds.}  
\end{figure}

For $p=2$ and $t\geq 1$ all polynomials $P(t)$ vanish
identically. This can be shown as follows. Let us consider the
probability $R(t)$ that a cluster dies out at time $t$, i.e., the
row at time $t$ is the last row reached by a cluster. Obviously
$R(t)$ is related to the survival probability by
\begin{equation}
\label{RDef}
R(t) = P(t)-P(t+1) \,.
\end{equation}
Clearly, $R(t)$ can be expressed as a weighted sum over the same set
of clusters as in~(\ref{SurvivalProbability}). However, in the present
case the weights differ from those in
Equation~(\ref{SurvivalProbability}) by the number of non-conducting
bonds in the clusters hull between $t$ and $t+1$  since it is now
required that all sites at time t+1 are inactive (see
Figure~\ref{FigCluster2}). This means that $R(t)$ can be expressed as
\begin{equation}
\label{DieoutProbability}
R(t) = \sum_{\cal C} p^n (1-p)^m (1-p)^{2k}\,,
\end{equation}
where $n$, $m$ and ${\cal C}$ have the same meaning as
in~(\ref{SurvivalProbability}) and $k$ is the number of active sites in the
horizontal row at time $t$. Obviously, for $p=2$ the additional factor
$(1-p)^{2k}$ drops out so that $P(t)=R(t)$ for all values of $t$. Moreover,
$R(0)=P(0)=1$ (for $p=2$). Combining these results with
Equation~(\ref{RDef}) we arrive at $P(t)=0$ for $t>0$, which completes the proof.
\subsection{Exact solution for $p$ at the Golden Ratio}
For $p=(1\pm \sqrt{5})/2$ we find that the survival probability `oscillates' between two different values, namely
\begin{equation}
\label{MainResult}
P(t) = 
\cases{
1 & if $t$ is even\\
\frac{\pm \sqrt{5}-1}{2} & if $t$ is odd.\\}
\end{equation}
\begin{figure}[b]
\epsfxsize=120mm
\centerline{\epsffile{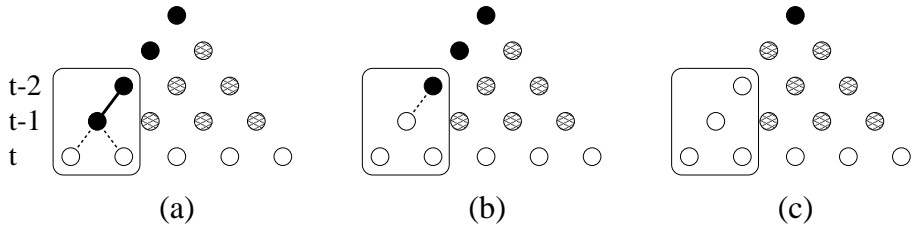}}
\caption{\label{FigProof}
Decomposition of the configurational sum of non-surviving
clusters into three subsets.
Open and closed bonds are denoted by solid (dashed) lines.
Bonds which are not shown may be either open or closed.
The box includes all bonds contributing to the factor $\hat{Q}_1$
while all other bonds contribute to the factor $\hat{Q}_2$. The
proof shows that the configurations in (a) and (b) cancel each other 
so that only the configurations of (c) contribute to $Q$. Iterating
the procedure by shifting the box to the right, it can be
shown that all sites at $t-2$ and $t-1$ have to be zero.}
\end{figure}
To prove this result, we first verify that~(\ref{MainResult}) is indeed
satisfied for $t=0$ and $t=1$. Then we show that 
\begin{equation}
P(t) = P(t-2) \qquad \mbox{for } t\geq 2 \quad \mbox{and}\quad p=(1\pm \sqrt{5})/2\,.
\end{equation}
However, instead of analyzing the survival probability directly, it
turns out to be more convenient to consider the complementary probability
$Q(t)=1-P(t)$ that a cluster does {\em not} survive until time
$t$. Obviously, $Q(t)$ is the sum over the weights of all clusters
which do not reach the horizontal row at time $t$, i.e., we impose the
boundary condition $s_{0,t}=s_{1,t}=\ldots=s_{t,t}=0$. Depending on
the states of the two sites $s_{0,t-1}$ and $s_{0,t-2}$ at the left
edge of the clusters, this set of clusters may be separated into three
different subsets, namely,\\

        {\bf(a)} a subset where $s_{0,t-1}=s_{0,t-2}=1$,

        {\bf(b)} a subset where $s_{0,t-1}=0$ and $s_{0,t-2}=1$, and

        {\bf(c)} a subset where $s_{0,t-1}=s_{0,t-2}=0$.\\

\noindent
Next we show that the weights of the clusters in the subsets (a) and (b) cancel each
other. To this end we note that the weighted sum $\hat{Q}(t)$ over all
clusters in subsets $(a)$ and $(b)$ may be decomposed into two independent factors
$\hat{Q}(t)=\hat{Q}_1\hat{Q}_2$, where $\hat{Q}_1$ depends only on the state of the three bonds
between the sites $s_{0,t-2}$, $s_{0,t-1}$, $s_{0,t}$, and $s_{1,t}$ (inside
the box in Figure~\ref{FigProof}), while $\hat{Q}_2$ accounts for all other relevant
bonds. Obviously, the first factor is given by
\begin{equation}
\hat{Q}_1^{(a)} = p(1-p)^2 \,, \qquad \hat{Q}_1^{(b)} = 1-p \,,
\end{equation}
while $\hat{Q}_2$ takes the same value in both subsets. Thus, if $p$ is
given by the Golden Ratio, we obtain $\hat{Q}_1^{(a)}+\hat{Q}_1^{(b)}=\hat{Q}_1(t)=0$
and therefore the weights of subsets (a) and (b) cancel each other. Consequently,
all remaining contributions to $Q(t)$ come from the clusters in subset
(c) where the sites $s_{0,t-1}$ and $s_{0,t-2}$ are inactive.
Now we can iterate this procedure by successively considering the
sites $s_{j,t-1}$ and $s_{j,t-2}$ from the left to the right, where
$j=1\ldots t-2$. In this way it can be shown  that all these sites
have to be inactive as well. Therefore, the only surviving
contributions are those in which the entire row of sites at $t-2$ is
inactive, implying that $Q(t)=Q(t-2)$. The proof of
Equation~(\ref{MainResult}) then follows by induction.
\section{Conclusions}\label{conclusion}
In this paper we have investigated the applicability of Yang-Lee
theory to the nonequilibrium phase transition of directed
percolation. In Section~\ref{intro} we developed an idea of how to
transfer the concepts of Yang and Lee to the phase transition of DP by
studying the complex zeros of the survival probability. In
Section~\ref{universality} we showed that a similar scenario as
observed for equilibrium phase transitions is also suitable for DP,
namely complex zeros approaching the critical point for increasing
system size. Moreover, we could extract the value of the critical
exponent for the temporal correlation length from the distribution of
zeros. Hence the zeros encode univerals properties of the phase
transition of DP. Section~\ref{Exact} presented exact results on
the survival probability of bond percolation which may be helpful for
further investigations. More precisely, we proved
that there are certain non-trivial values of $p$ for which the values
of the polynomials $P(t)$ can be calculated exactly for all times.       
\ack
This work was supported by the DAAD/CAPES within the German-Brazilian
cooperation project PROBRAL -- ``Rigorous Results in Nonequilibrium
Statistical Mechanics and Nonlinear Physics''.
\appendix
\section{MC-Simulations}\label{MC}
Here we demonstrate that the value of the survival probability for a
complex percolation probability $p$ as well as the polynomials $P(t)$
are accessible by computer simulations. 

Let us consider the definition of the
survival probability as a sum over configurations of surviving
clusters in Equation~(\ref{SurvivalProbability}). Formally this
expression can be rewritten as
\begin{equation}
\label{ComplexMC}
P(t) = \sum_{\cal C} p^n (1-p)^m =
\sum_{\cal C} q^n (1-q)^m \, 
\underbrace{\frac{p^n (1-p)^m }{q^n (1-q)^m }}_{f_{n,m}}
\end{equation}
where $n$ and $m$ denote the number of bonds of the cluster and of
its hull respectively. Therefore, instead of simulating the system
at a given $p$, we may simulate it using a {\em different}
percolation probability $q$ reweighting each cluster by the factor
$f_{n,m}$. While $q \in (0,1)$ still has to be a real number, $p$ is
no longer restricted to be real, it can be any complex number. Using
this reweighting technique it is in principle possible to access the
entire complex plane by numerical simulations. For any finite $t$ such
a simulation is stable and will converge to the
correct result. But even for small $t$ the convergence time can be very 
long, limiting the range of applications.

Using the reweighting technique it is also possible to approximate
the coefficients of the polynomial $P(t) = \sum_k a_k p^k$ by
considering $p$ as a free parameter and expanding the term $(1-p)^m$
in Equation~(\ref{ComplexMC}). The coefficients are then given by
\begin{equation}
a_k = \sum_{\cal C} \, q^n(1-q)^m \, \frac{C(m,k-n) (-1)^{k-n}}{q^n (1-q)^m}\,,
\end{equation}
where $q \in (0,1)$ is again a free parameter and
\begin{equation}
C(m,k-n) = \left\{
\begin{array}{cc}
\frac{m!}{(k-n)!(m-k-n)!} & \mbox{ if } 0 \leq k-n \leq m \\
0 & \mbox{ otherwise }. 
\end{array}\right.
\end{equation}
However, in most cases the direct construction of the polynomials
using symbolic algebra turns out to be more efficient.
\section{First coefficients of the polynomials $P(t)$ in the limit $t \to \infty$}\label{coefficients}
As can be seen in Equation~(\ref{polynomials}) the first non-vanishing
coefficient of the polynomial for the survival probability
$P(t)=\sum_na_np^n$ for DPb is always a power of $2$. This contribution
corresponds to the configurational weight of a surviving path without
branches. The purpose of this appendix is to point out that even the
following coefficients can be computed exactly, provived that $t$ is
large enough. More specifically, we conjecture that
\begin{equation}
a_{t+k} = 2^{t-k-1} \, q_k(t)\,,
\end{equation}
where 
\begin{equation}
q_k(t) = \sum_{m=0}^M b_m^{(k)} t^m \,,\qquad 
M=
\left\{
\begin{array}{ll} 
k/2 & \mbox{ if } k \mbox{ even} \\
(k-1)/2 & \mbox{ if } k \mbox{ odd}
\end{array}
\right.
\end{equation}
is a finite polynomial in $t$. Using the explicit expressions for $P(t)$ up to $t=15$ we
find that the first eight polynomials read
\footnotesize
\begin{eqnarray}
q_0(t) &=& \nonumber  2 \\
q_1(t) &=& \nonumber  -2 \\
q_2(t) &=& \nonumber  -4-2t \\
q_3(t) &=& \nonumber  -2t \\
q_4(t) &=&   -16+3t+t^2 \\
q_5(t) &=& \nonumber  32+t+3t^2 \\
q_6(t) &=& \nonumber  -232 + \frac{64}{3}t+3t^2-\frac{1}{3}t^3 \\
q_7(t) &=& \nonumber  808 - \frac{34}{3} + 5t^2 - \frac{5}{3}t^3 
\end{eqnarray}
\normalsize
We conjecture that the leading coefficient of these
polynomials is given by
\begin{equation}
b_M^{(k)} = \left\{
\begin{array}{ll}
2 (-1)^M/M! & \mbox{ if } k \mbox{ even } \\
2 (k-1)(-1)^M/M! & \mbox{ if } k \mbox{ odd }. 
\end{array}
\right.
\end{equation}
This implies that the first non-vanishing coefficients of the
polynomial $P(t)$ grow in such a way that their limit
$\lim_{t\to\infty} \frac{a_{t+k}}{2^tt^M}$ for fixed $k$ is well-defined. Summing up these
contributions we obtain for the survival probability
\begin{eqnarray}
\label{AppResult}
P(t) &\simeq& 
(2p)^t\sum_{j=0}^{\infty} \frac{(-p^2t/4)^j}{j!}\,\Bigl(1+p(j-1/2)\Bigr) \\
&=& \frac{1}{4}(2p)^te^{-p^2t/4} (4-2p-p^3t)\ . \nonumber
\end{eqnarray}
Physically this expression corresponds to loop-free graphs and serves
as an approximation close
to $p=0$. The convergence radius of Equation~(\ref{AppResult}) determined
from $2p=e^{p^2/4}$ is $|p|<0.53744$. Note that none of the
non-trivial roots computed up to $t=15$ lies inside this
radius. We conjecture that this might be true for any $t$.



\section*{References}
\begin{thebibliography}{99}

\bibitem{YangLee} Yang~C~N and Lee~T~D (1952) {\it Phys. Rev.} {\bf 78} 404;
Lee~T~D and Yang~C~N (1952) {\it Phys. Rev.} {\bf 87} 410

\bibitem{Arndt} Arndt~P~F (2000) {\it Phys. Rev. Lett.} {\bf 84} 814

\bibitem{Physica} Arndt~P~F, Dahmen~S~R and Hinrichsen~H (2001)
{\it Physica} A {\bf 295} 128

\bibitem{YangLeearticles} Alves~N~A, Drugowich~de~Felicio~J~R and
  Hansmann~U~H~E (2000) {\it J. Phys.} A {\bf 33} 7489; Alves~N~A, Drugowich~de~Felicio~J~R and
  Hansmann~U~H~E (1997) {\it Inter. J. Mod. Phys.} C {\bf 8} 1063

\bibitem{Kinzel} Kinzel~W (1983) {\it Annals of the Israel Physical Society} 
vol~5, ed. by Deutscher~G, Zallen~R and Adler~J (Bristol: Adam Hilger)

\bibitem{MarroDickman} Marro~J and Dickman~R (1999) {\it Nonequilibrium phase
transitions in lattice models} (Cambridge: Cambridge University Press)

\bibitem{Hinrichsen1} Hinrichsen~H (2000) {\it Adv. Phys.} {\bf 49} 815

\bibitem{Hinrichsen2} Hinrichsen~H (2000) {\it Braz. J. Phys.} {\bf 30} 69

\bibitem{Ziff} Voigt~C~A and Ziff~R~M (1997) {\it Phys. Rev.} E {\bf 56} R6241

\bibitem{Bayreuth} Rupp~P, Richter~R and Rehberg~I, cond-mat/0201308

\bibitem{Jensen} Jensen~I (1996) \PRL {\bf 77} 4988

\bibitem{Yu} Tretyakov~A~Y and Inui~N (1995) {\it J. Phys.} A {\bf 28} 3985
  
\bibitem{henkelandschuetz} Henkel~M and Sch\"utz~G (1988) {\it
    J. Phys.} A {\bf 21} 2617

\bibitem{YangLeeNeuer} Derrida~B, De~Seze~L and Itzykson C (1983) {\it J. Stat. Phys.} {\bf 33} 559

\bibitem{NumberTheory} see e.g.:
Schroeder~M (1999) {\it Number Theory in Science and Communication} (Springer
Series in Information Science, vol.~7, Springer Verlag); Vajda~S (1989) {\it Fibonacci and Lucas numbers, and the Golden
  Section} (Chichester: Horwood)

\bibitem{chaotic} Ozorio~de~Almeida~A~M (1988) {\it Hamiltonian Systems: Chaos
    and Quantization} (Cambridge: Cambridge University Press)

\bibitem{arts} Ghyka~M (1977) {\it The Geometry of Art and Life} (New York: Dover Publications)

\end{thebibliography}
\end{document}